\documentclass[a4paper,12pt]{article}
\pdfoutput=1 
\newcommand{\nc}{\newcommand*} 

\nc{\al}{\alpha}
\nc{\s}{\sigma}
\nc{\dt}{\delta}
\nc{\Dt}{\Delta}
\nc{\Ld}{\Lambda}
\nc{\p}{\partial}
\nc{\Om}{\Omega}
\nc{\rd}{\mathrm{d}}
\nc{\Od}{\mathcal{O}} 

\def\({\left(}
\def\){\right)}
\def\[{\left[}
\def\]{\right]}
\def\e{\begin{equation}}
\def\q{\end{equation}}
\def\m{\begin{eqnarray}}
\def\n{\end{eqnarray}}

\nc{\Eq}[1]{Eq.~\eqref{#1}}     
\nc{\Fig}[1]{Fig.~\ref{#1}}     
\nc{\Table}[1]{Table~\ref{#1}}  
\nc{\Sec}[1]{Sec.~\ref{#1}}     

\nc{\Msun}{M_\odot}             
\nc{\fpbh}{f_{\mathrm{pbh}}}    
\nc{\fpbhn}{f_{\mathrm{PBH0}}}    
\nc{\mR}{\mathcal{R}} 
\nc{\seq}{\sigma_{\mathrm{eq}}}
\nc{\ogw}{\Omega_{\mathrm{GW}}}
\nc{\gpcyr}{\mathrm{Gpc}^{-3}\,\mathrm{yr}^{-1}}
\nc{\lvc}{LIGO/Virgo} 
\nc{\SNR}{\mathrm{SNR}} 
\nc{\mmin}{{m_{\mathrm{min}}}}
\nc{\mmax}{{m_{\mathrm{max}}}}
\nc{\Mmin}{{M_{\mathrm{min}}}}
\nc{\fmin}{{f_{\mathrm{min}}}}
\nc{\VT}{\mathrm{VT}}
\nc{\rhoGW}{\rho_{\mathrm{GW}}}
\nc{\vth}{\vec{\theta}}
\nc{\vd}{\vec{d}}
\nc{\vla}{\vec{\lambda}}
\nc{\Nobs}{N_{\mathrm{obs}}}
\nc{\av}[1]{\langle #1 \rangle} 
\nc{\km}{\mathrm{km}}
\nc{\Mpc}{\mathrm{Mpc}}
\nc{\Tobs}{T_{\mathrm{obs}}}
\nc{\Ntemp}{N_{\mathrm{temp}}}
\nc{\cH}{{\mathcal{H}}}

\nc{\OL}{{\Omega_\Lambda}}
\nc{\OM}{{\Omega_\mathrm{m}}}
\nc{\OR}{{\Omega_r}}
\nc{\mU}{{\mathcal{U}}}
\nc{\Mc}{{M_\mathrm{c}}}
\nc{\sgc}{{\sigma_\mathrm{c}}}
\nc{\Mf}{{M_\mathrm{f}}}
\nc{\LCDM}{{$\Lambda$CDM}}
\nc{\kmsmpc}{km~s$^{-1}$~Mpc$^{-1}$}

\nc{\addref}{[\textcolor{red}{add ref}] } 
\nc{\eg}{\textit{e.g.~}}
\nc{\app}{\approx}
\nc{\hf}{\frac{1}{2}}
\nc{\discuss}{\textcolor{red}{Add discussion here!}}
\nc{\red}[1]{\textcolor{red}{#1}}

\usepackage{jcappub} 
\DeclareUnicodeCharacter{2212}{\ensuremath{-}}
\usepackage[T1]{fontenc} 
\usepackage{bm}
\usepackage{indentfirst}
\usepackage{amsmath}
\usepackage{graphicx}
\usepackage{float}
\usepackage{amssymb}
\usepackage{subfigure}
\usepackage{amssymb}
\usepackage{psfrag}
\usepackage[
colorlinks=true,        
citecolor=blue,         
linkcolor=blue,         
urlcolor=blue           
]{hyperref}             
\usepackage{epstopdf}
\usepackage{diagbox}
\usepackage{tensor}     
\usepackage{float}
\usepackage{bm}         
\usepackage{xcolor}     
\usepackage{lipsum}
\usepackage{color}      
\usepackage[utf8]{inputenc} 
\usepackage{adjustbox}
\usepackage[section]{placeins} 
\usepackage{multirow}

\title{Constraints on Primordial-Black-Hole Population and Cosmic Expansion History from GWTC-3}

\author{Zu-Cheng Chen$^{1, 2}$}
\author{Shen-Shi Du$^{3}$}
\author{Qing-Guo Huang$^{4, 5, 6}$}
\author{Zhi-Qiang You$^{*1, 2}$}

\affiliation{$^1$Department of Astronomy, Beijing Normal University, Beijing 100875, China}
\affiliation{$^2$Advanced Institute of Natural Sciences, Beijing Normal University, Zhuhai 519087, China}
\affiliation{$^3$School of Physics and Technology, Wuhan University, Wuhan, Hubei 430072, China}
\affiliation{$^4$CAS Key Laboratory of Theoretical Physics, Institute of Theoretical Physics, Chinese Academy of Sciences, Beijing 100190, China}
\affiliation{$^5$School of Physical Sciences, University of Chinese Academy of Sciences, No. 19A Yuquan Road, Beijing 100049, China}
\affiliation{$^6$School of Fundamental Physics and Mathematical Sciences, Hangzhou Institute for Advanced Study, UCAS, Hangzhou 310024, China}

\emailAdd{zucheng.chen@bnu.edu.cn}
\emailAdd{$\,\,\,\,\,$huangqg@itp.ac.cn}
\emailAdd{$\,\,\,\,\,\,\,\,\,\,\,\,\,\,\,\,\,\,\,\,$Corresponding author: zhiqiang.you@bnu.edu.cn}

\abstract{Gravitational waves (GWs) from compact binary coalescences provide an independent probe of the cosmic expansion history other than electromagnetic waves. In this work, we assume the binary black holes (BBHs) detected by LIGO-Virgo-KAGRA (LVK) collaborations are of primordial origin and constrain the population parameters of primordial black holes (PBHs) and Hubble parameter $H(z)$ using $42$ BBHs from third LVK GW transient catalog (GWTC-3). Three PBH mass models are considered: lognormal, power-law, and critical collapse PBH mass functions. By performing a hierarchical Bayesian population analysis, the Bayes factor strongly disfavors the power-law PBH mass function against the other two in GWTC-3. The constraints on standard $\Lambda{\rm CDM}$ cosmological parameters are rather weak and in agreement with current results. When combining the multi-messenger standard siren measurement from GW170817, the Hubble constant $H_0$ is constrained to be $69^{+19}_{-8}\, \mathrm{km}\, \mathrm{s}^{-1}\, \mathrm{Mpc}^{-1}$ and $70^{+26}_{-8}\, \mathrm{km}\, \mathrm{s}^{-1}\, \mathrm{Mpc}^{-1}$ at $68\%$ confidence for the lognormal and critical collapse mass models, respectively.
Furthermore, we consider a mixed ABH+PBH model, in which we assume LVK BBHs can come from both the astrophysical black hole (ABH) and PBH channels. We find that the ABH+PBH model can better describe the mass distribution in GWTC-3 than any single ABH or PBH model, thus improving the precision to constrain the Hubble constant. With the increased BBH events, the mixed ABH+PBH model provides a robust statistical inference for both the population and cosmological parameters.
}

\begin{document}
	\maketitle
	\flushbottom
	
	\section{\label{intro}Introduction}
	The Hubble parameter $H(z)$ is a fundamental observable in probing the cosmic expansion history and elucidating the nature of the dark energy component. The Planck \cite{Planck:2018vyg} cosmic microwave background observations provide the hitherto most precise measurement of its present value, Hubble constant, as $H_0 = 67.36 \pm 0.54$ \kmsmpc, based on the $\Lambda$ Cold Dark Matter ($\Lambda$CDM) cosmological model. Meanwhile, the recent local measurement by the SH0ES team using the Cepheid variable calibrated Type Ia supernovae gives $H_0=73.30\pm1.04$ \kmsmpc \cite{Riess:2021jrx}. It cracks at $>$ 5$\sigma$ level for the early and late Universe results, known as the Hubble tension, spurring intense debate of either new physics beyond $\Lambda$CDM or unaccounted-for systematics in current observations. Although both the early and late Universe solutions have been extensively investigated (see \eg recent reviews \cite{Shah:2021onj,DiValentino:2021izs,Perivolaropoulos:2021jda}), this tension remains to be solved.

Gravitational waves (GWs) are new independent probes that have the potential to understand the inconsistency between the various measurements.
GWs produced by compact binary coalescences can be standard sirens in the context of the general theory of relativity -- the strain amplitude encodes an absolute distance to the source  \cite{Schutz:1986gp,Holz:2005df}.
Combined with the redshift informed by electromagnetic (EM) counterparts, GW sirens provide a novel estimate of $H_0$ without using a distance ladder. This standard siren has been proven promising by the measurement with the joint GW-EM detections of binary neutron star merger event GW170817 \cite{LIGOScientific:2017adf}.

Even in the absence of EM observations, GWs alone can probe the cosmic expansion history if the cosmological parameters are analyzed simultaneously with the population parameters of compact binaries. It can be achieved because cosmology determines how the observed (redshifted) masses scale with luminosity distance.
In this sense, we can infer the cosmic expansion history without resorting to the cross-correlation technique \cite{Mukherjee:2020hyn,Diaz:2021pem,Mukherjee:2022afz}.
In fact, this method has been applied to the third LIGO-Virgo-KAGRA (LVK) GW transient catalog (GWTC-3)  \cite{LIGOScientific:2021djp} assuming that the masses and redshift of binary black holes (BBHs) follow some phenomenological distributions inspired by the astrophysical black hole (ABH) model \cite{LIGOScientific:2021aug,Karathanasis:2022rtr}, as well as in the prediction of Hubble constant for the third-generation observatories like the Einstein Telescope and Cosmic Explorer \cite{You:2020wju}.

It has been speculated that the LVK BBHs are from primordial black holes (PBHs) \cite{Sasaki:2016jop} since the detection of the first GW event, GW150914 \cite{LIGOScientific:2016aoc}, as the mass of BBHs observed by GWs is unexpected heavier than those observed by X-rays \cite{Wiktorowicz:2013dua,Casares:2013tpa,Corral-Santana:2013uua}.
PBHs are black holes formed in the very early universe due to the collapse of primordial density perturbations \cite{Hawking:1971ei,Carr:1974nx}. They can not only explain LVK BBHs \cite{Chen:2018czv,Chen:2018rzo,Chen:2019irf,Wu:2020drm,Chen:2021nxo,Liu:2022iuf,Zheng:2022wqo}, but also serve as cold dark matter (CDM) candidates. Moreover, the population properties of PBH binaries are quite different from those of ABH binaries. For instance, the merger rate of PBH binaries grows as the redshift $z$ increases, while the merger rate of ABH binaries peaks at $z\sim2$ and then decreases rapidly. This feature can be used to distinguish PBHs from ABHs \cite{Chen:2019irf,Mukherjee:2021ags}. Also, multiband GW observations can help test PBH and measure the Hubble constant \cite{Liu:2021jnw}.

In this work, we simultaneously infer the cosmic expansion history with the BBH population properties under the PBH scenario using GWTC-3 data release.
The remainder of this paper is organized as follows. In \Sec{PBH}, we consider three different PBH mass spectra and review the merger rate density distribution of PBH binaries. In \Sec{ABHPBH}, we introduce a mixed model which contains BBHs from both the ABH and PBH channels. In \Sec{Bayes}, we describe the hierarchical Bayesian inference used to infer the model parameters.
In \Sec{results}, we present the constraints on the Hubble constant, the PBH and ABH+PBH population properties.
Finally, we present our conclusions in \Sec{conclusion}. Additionally, we put full posteriors for the three models with different PBH mass spectra and the ABH+PBH scenario in the Appendix.

\section{\label{PBH}PBH Scenario}	
In order to infer the PBH population parameters and cosmological parameters from the GWTC-3, one needs to work out the merger rate density distribution of PBH binaries. This section briefly reviews the PBH scenario under the assumption that PBH binaries are formed in the early Universe and are effectively randomly distributed in space \cite{Ali-Haimoud:2017rtz}.
Two neighboring PBHs will decouple from the background of the expanding Universe due to their gravitational attraction as long as they are close enough. The decoupling from the Hubble flow usually happens deep in the radiation dominated era \cite{Ali-Haimoud:2017rtz}. The tidal force from other PBHs and matter density perturbations will provide an angular momentum to this pair of PBHs, preventing them from direct coalescence. This PBH pair will therefore form a binary. After the formation of a binary, the orbit of this system will shrink due to the GW radiation. The PBH binaries will eventually merge and potentially be detected by GW detectors, thus explaining LVK BBH events.

Given an extended PBH mass function, the merger rate density distribution for the PBH binaries has been worked out in \cite{Chen:2018czv} by accounting for the torques from all PBHs and linear density perturbations. The redshift-dependent comoving merger rate density in units of $\Msun^{-2}\,\gpcyr$ takes the following form \cite{Chen:2018czv}
\begin{equation}
	\label{calR}
	\begin{split}
		\mR(m_1, m_2, z) &\app 2.8 \cdot 10^6  \({\frac{t(z)}{t_0}}\)^{-\frac{34}{37}} \fpbh^2 (0.7\fpbh^2+\sigma_{\mathrm{eq}}^2)^{-{21\over 74}} \\
		& \times  \min\(\frac{P(m_1)}{m_1}, \frac{P(m_2)}{m_2}\) \({P(m_1)\over m_1}+{P(m_2)\over m_2}\)\\
		& \times (m_1 m_2)^{{3\over 37}} (m_1+m_2)^{36\over 37},
	\end{split}
\end{equation}
where the component masses $m_1$ and $m_2$ are in units of $\Msun$, $t(z)$ is the cosmic time at redshift $z$, and $t_0 \equiv t(0)$.
Here, $\fpbh$ is the abundance of PBH in CDM, and $\sigma_{\mathrm{eq}}^2$ is the variance of density perturbations of the rest of dark matter at radiation-matter equality, with $\sigma_{\mathrm{eq}} \approx 0.005$ \cite{Ali-Haimoud:2017rtz}.
In this work, we use the units in which the speed of light $c=1$.
Note that the redshift evolution of the merger rate of PBH binaries follows a power-law form as $(t(z)/t_0)^{-34/37}$, which is quite different from that of astrophysical black hole (ABH) binaries and can be used to distinguish between PBHs and ABHs \cite{Chen:2019irf}.
In \Eq{calR}, the PBH mass function $P(m)$ has been normalized to unity, namely
\e
\int_0^\infty P(m) dm = 1.
\q
In the following, we will consider three types of PBH mass distributions originating from different PBH formation models.

The first one is the lognormal mass function taking the form of \citep{Dolgov:1992pu}
\e\label{log}
P(m, \sigma_{\mathrm{c}}, \Mc) = \frac{1}{\sqrt{2\pi} \sigma_{\mathrm{c}} m} \exp \(-\frac{\ln^2\(m/\Mc\)}{2\sigma_{\mathrm{c}}^2}\),
\q
where $\Mc$ is the peak mass of $m P(m)$, and $\sigma_{\mathrm{c}}$ gives the width of the mass spectrum.
The lognormal mass function is often a good approximation to a large class of extended mass distributions if PBHs are formed from a smooth symmetric peak in the inflationary power spectrum when the slow-roll approximation holds \cite{Carr:2017jsz,Green:2016xgy,Kannike:2017bxn}.

The second one is the power-law mass function of the form \cite{Carr:1975qj,DeLuca:2020ioi}
\e
P(m, M_{\min}) = \hf M_{\min}^{1/2}\, m^{-3/2}\, \Theta(m - M_{\min}),
\q
where $M_{\min}$ is the lower mass cut-off of the mass spectrum. The power-law mass function typically results from a broad or flat power spectrum of the curvature perturbations \cite{DeLuca:2020ioi} during radiation dominated era \cite{Carr:2017jsz}.

The third one is the critical collapse mass function taking the form of \cite{Niemeyer:1997mt,Yokoyama:1998xd,Carr:2016hva,Gow:2020cou}
\e
P(m, \alpha, M_{\mathrm{f}})=\frac{\alpha^2\,  m^\alpha}{M_{\mathrm{f}}^{1+\alpha}\, \Gamma(1 / \alpha)} \exp \left(-(m/M_{\mathrm{f}})^{\alpha}\right),
\q
where $\alpha$ is a universal exponent which is related to the critical collapse of radiation, and $M_{\mathrm{f}}$ is a mass scale at the order of horizon mass at the collapse epoch \cite{Carr:2016hva}. This mass function is supposed to be closely associated with a monochromatic power spectrum of the density fluctuations. In this case, there is an exponential upper cut-off at a mass scale of $M_{\mathrm{f}}$, but no lower mass cut-off. Here and after, we dub it as CC mass function.

\section{\label{ABHPBH}ABH+PBH Scenario}

Previous analyses from GWTC-2 indicate that the LVK BBHs may comprise both the ABHs and PBHs \cite{Hall:2020daa,Hutsi:2020sol,Wong:2020yig,DeLuca:2021wjr,Franciolini:2021tla}. In this section, we consider a mixed ABH+PBH model in which BBHs can come from both the ABH and PBH channels.

We take the lognormal mass function for the PBH part and a phenomenological model following Ref.~\cite{LIGOScientific:2021aug} for the ABH part.
The mixed merger rate is a summation of the ABH merger rate and the PBH merger rate, namely,
\begin{equation}
	\mR_{\mathrm{total}}(m_1, m_2, z) = \mR_{\mathrm{ABH}}(m_1, m_2, z) + \mR_{\mathrm{PBH}}(m_1, m_2, z),
\end{equation}
where $\mR_{\mathrm{PBH}}(m_1, m_2, z)$ is given by \Eq{calR}. For the PBH model, we adopt the lognormal mass function widely used in the literature.
For the ABH model, we model the binary merger rate using a phenomenological model following Ref.~\cite{LIGOScientific:2021aug}. To be specific, the ABH merger rate is estimated as
\begin{equation}
	\mR_{\mathrm{ABH}}(m_1, m_2, z) = R_{0, \mathrm{ABH}} \, \pi(z) \, \pi(m_1)\, \pi(m_2),
\end{equation}
where $R_{0, \mathrm{ABH}}$ is the local merger rate of ABH binaries. We parameterize the redshift distribution $\pi(z)$ as
\begin{equation}
	\pi\left(z| \gamma, \kappa, z_{\mathrm{p}}\right)=\left[1+\left(1+z_{\mathrm{p}}\right)^{-\gamma-k}\right]  \frac{(1+z)^\gamma}{1+\left[(1+z) /\left(1+z_{\mathrm{p})}\right]^{\gamma+k}\right.},
\end{equation}
where $\gamma$ and $k$ are the slopes of the two power-law regimes before and after a turning point $z_{\mathrm{p}}$. This parameterization is motivated by the fact that the binary formation rate might follow the star formation rate \cite{Madau:2014bja,Callister:2020arv}.
The primary mass distribution $\pi(m_1)\equiv\pi\left(m_1| m_{\min }, m_{\max }, \alpha, \lambda_{\mathrm{g}}, \mu_{\mathrm{g}}, \sigma_{\mathrm{g}}\right)$ is composed of a power-law and Gaussian component, namely
\begin{eqnarray}	
	\pi(m_1)=\left[(1-\lambda_{\mathrm{g}}) \mathcal{P} (m_1| m_{\min }, m_{\max },-\alpha)+\lambda_{\mathrm{g}} \mathcal{G}(m_1 | \mu_{\mathrm{g}}, \sigma_{\mathrm{g}})\right] S(m_1, m_{\min }, \delta_m),
\end{eqnarray}
where $\mathcal{P}(x | x_{\min }, x_{\max }, \alpha)$ is a truncated power law described by slope $\alpha$, and lower and upper bounds $x_{\min }, x_{\max }$ at which there is a hard cut-off,
\begin{equation}
	\mathcal{P}(x | x_{\min }, x_{\max }, \alpha) \propto \begin{cases}x^\alpha & \left(x_{\min } \leqslant x \leqslant x_{\max }\right) \\ 0 & \text { otherwise. }\end{cases}
\end{equation}
The function $\mathcal{G}(x | \mu, \sigma, a, b)$ is a Gaussian distribution with mean $\mu$ and standard deviation $\sigma$,
\begin{equation}
	\mathcal{G}(x | \mu, \sigma, a, b)=\frac{1}{\sigma \sqrt{2 \pi}} \exp \left[-\frac{(x-\mu)^2}{2 \sigma^2}\right].
\end{equation}
Meanwhile, $\lambda_{\mathrm{g}}$ is a ratio parameters of these two component $\mathcal{P}$ and $\mathcal{G}$. The function $S(m_1, m_{\text {min }}, \delta_m)$ is a sigmoid-like window function that indicates a smoothing rise in the interval $\left(m_{\min }, m_{\min }+\delta_m\right)$:
\begin{equation}
	S(m, m_{\min }, \delta_m)= \begin{cases}0 & \left(m<m_{\min }\right) \\ {\left[f\left(m-m_{\min }, \delta_m\right)+1\right]^{-1}} & \left(m_{\min } \leq m<m_{\min }+\delta_m\right) \\ 1 & \left(m \geq m_{\min }+\delta_m\right),\end{cases}
\end{equation}
with
\begin{equation}
	f(m^{\prime}, \delta_m)=\exp \left(\frac{\delta_m}{m^{\prime}}+\frac{\delta_m}{m^{\prime}-\delta_m}\right).
\end{equation}
The secondary mass distribution $\pi(m_2)\equiv\pi\left(m_2 \mid m_1, m_{\min }, \alpha\right)$ is described with a truncated power-law with slope $\beta$ between a minimum mass $m_{\min }$ and a maximum mass $m_1$,
\begin{equation}
	\pi(m_2 | m_1, m_{\min }, \alpha)=\mathcal{P}(m_2 \mid m_{\min }, m_1, \beta)\, S(m_2, m_{\min }, \delta_m).
\end{equation}

\section{\label{Bayes}Data and Methodology}

In this work, we use BBH events from the GWTC-3 \cite{LIGOScientific:2021djp} to jointly infer the PBH population parameters and cosmological parameters. GWTC-3 contains 90 GW candidates detected during the first three LVK observing runs.
Following \cite{LIGOScientific:2021aug}, we use $42$ BBH candidates with network-matched filter signal-to-noise ratio larger than $11$ and inverse false alarm rate higher than $4$ year. A summary of their properties can be found in Table~1 of \cite{LIGOScientific:2021aug}.
In the analyses, we use combined posterior samples obtained from the IMRPhenom \cite{Thompson:2020nei,Pratten:2020ceb} and SEOBNR \cite{Ossokine:2020kjp,Matas:2020wab} waveform families.

For each BBH event, GW experiments measure the luminosity distance $D_{\mathrm{L}}$ and redshifted masses $m_{1}^{\mathrm{det}}, m_{2}^{\mathrm{det}}$, instead of the redshift $z$ and source masses $m_{1}$, $m_{2}$.
These quantities are related by
\e\label{mz}
m_{i}=\frac{m_{i}^{\text {det }}}{1+z\left(D_{\mathrm{L}} ; H_{0}, \Omega_{\mathrm{m}}\right)},
\q
where $H_0$ is the Hubble constant, and $\Omega_{\mathrm{m}}$ is the present-day matter density.
The relation \Eq{mz} can be used to probe cosmic expansion history even without resorting to the redshift information inferred from electromagnetic counterpart \cite{Taylor:2011fs,Taylor:2012db} provided source mass distribution can be well characterized.

We consider the flat $\Lambda$CDM Universe in this work. The Hubble rate at redshift $z$ is
\e
H(z) = H_0 E(z),
\q
where $H_0 \equiv h \times 100 \mathrm{~km} \mathrm{~s}^{-1} \mathrm{Mpc}^{-1}$ is the Hubble constant, and
\e
E(z) = \sqrt{\OM (1+z)^3 + (1-\OM)}.
\q
In the above equation, we have neglected the contribution from the radiation and neutrinos as we are interested in a small redshift range.
Given redshift $z$, one can then calculate the cosmic time $t$ as
\e
t(z) = \frac{1}{H_0} \int_{z}^{\infty} \frac{dz'}{E(z') (1+z')},
\q
and the luminosity distance $D_L$ as
\e\label{dl}
D_L(z) = \frac{(1+z)}{H_0} \int_{0}^{z} \frac{dz'}{E(z')}.
\q
Solving the above equation yields the redshift as a function of luminosity distance, $z(D_L)$, that is needed in \Eq{mz}.

We use the hierarchical Bayesian inference to infer the model parameters. To do so, we rewrite the merger rate density \Eq{calR} as
\e\label{calR2}
 \mR(\theta|\Phi) = R_0\, p(\theta|\Phi),
\q
where $\theta = \{m_1, m_2, z\}$ are the intrinsic GW parameters that are interesting for cosmology and unique for each event, while $\Phi$ denotes the hyper-parameters that are common to the entire population of GW sources. Concretely, $\Phi = \{H_0, \OM, \sigma_{\mathrm{c}}, \Mc\}$, $\{H_0, \OM, M_{\min}\}$, and $\{H_0, \OM, \alpha, M_{\mathrm{f}}\}$ for the cases of lognormal, power-law and CC mass function, respectively. The local merger rate $R_0$ in \Eq{calR2} is defined by
\e\label{R0}
R_0 = \int_{0}^{\infty}\int_{0}^{\infty} \mR(m_1, m_2, z=0|\Phi) dm_1 dm_2,
\q
ensuring that $p(\theta|\Phi)$ is normalized to unity when $z=0$. Given $R_0$ and $\Phi$, one can solve for $\fpbh$ using \Eq{calR} and \Eq{R0}.
Note that $p(\theta|\Phi)$ are measured in source frame, and can be converted to the detector frame by
\e\label{ppop}
p_{\mathrm{pop}}(\theta|\Phi) = \frac{1}{1+z} \frac{dV_\mathrm{c}}{dz} p(\theta|\Phi),
\q
where $dV_\mathrm{c}/dz$ is the differential comoving volume, and the factor $1/(1 + z)$ converts time increments from the source frame to the detector frame.

Given the data, $\textbf{d} = \{d_1, d_2, \cdots, d_{\mathrm{obs}}\}$, of $N_{\mathrm{obs}}$ GW events, we model the total number of events as an inhomogeneous Poisson process, yielding the likelihood \cite{Loredo:2004nn,Thrane:2018qnx,Mandel:2018mve}
\begin{equation}
		\mathcal{L}(\textbf{d}|R_0,\Phi) \propto R_0^{N_{\mathrm{obs}}} e^{-R_0 \xi(\Phi)} \prod_{i=1}^{N_{\mathrm{obs}}} \int \mathcal{L} (d_{i}| \theta)\, p_{\mathrm{pop}}(\theta|\Phi) d \theta,
\end{equation}
where $\mathcal{L} (d_{i}| \theta)$ is the individual likelihood for $i$th GW event that can be derived from the individual posterior by reweighing with the prior on $\theta$, and
\e\label{xi}
\xi(\Phi) = \int P_{\mathrm{det}}(\theta)\, p_{\mathrm{pop}}(\theta|\Phi)\, \mathrm{d} \theta
\q
is the detection fraction that quantifies selection biases for a population with parameters $\Phi$.
Here, $P_{\text{det}}(\theta)$ is the detection probability that depends on the source parameters $\theta$.
We use the simulated signals (injections) that are available in  \cite{ligo_scientific_collaboration_and_virgo_2021_5546676} to estimate the detection fraction.
In practice, \Eq{xi} is approximated by using a Monte Carlo integral over found injections \cite{gwtc3_mass_fun}
\begin{equation}
	\xi(\Phi) \approx \frac{1}{N_{\mathrm{inj}}} \sum_{j=1}^{N_{\text {found }}} \frac{p_{\mathrm{pop}}(\theta_{j} | \Phi)}{p_{\mathrm{draw}}(\theta_j)},
\end{equation}
where $N_{\text{inj}}$ is the total number of injections, $N_{\text{found}}$ is the number of injections that are successfully detected, and $p_{\text {draw }}$ is the probability distribution from which the injections are drawn.
We incorporate the PBH population distribution \eqref{ppop} into the \texttt{ICAROGW} \cite{Mastrogiovanni:2021wsd} package to estimate the likelihood function, and use \texttt{dynesty} \cite{Speagle:2019ivv} sampler called from \texttt{Bilby} \cite{Ashton:2018jfp,Romero-Shaw:2020owr} to search over the parameter space.

\section{\label{results}Results}

We use $42$ detected BBHs with SNR $>11$ to estimate the cosmological and population parameters. Similar to \cite{LIGOScientific:2021aug}, we consider two cosmological models: (i) a general \LCDM\ model with wide priors on the Hubble constant $H_0$ and matter density $\OM$, and (ii) a $H_0$-tension model with a fixed value of $\OM=0.315$ \cite{Planck:2018vyg} and with a restricted prior in the $H_{0}$ tension region ($H_0 \in[65,77]$ \kmsmpc). \Table{tab:priorss} summarizes the model parameters and their prior distributions used in the Bayesian parameter estimations. The prior ranges that we model are wide enough to include the effect of a possible time delay between the formation and the merger of the binary. The full posteriors for the (lognormal, power-law, and CC) PBH models  and the mixed ABH+PBH model considered in this work are presented in Appendix~\ref{post}.
\begin{table*}[htbp!]
	\centering
	\begin{adjustbox}{width=1\textwidth}
		\begin{tabular}{lll}
			\hline\hline
			\textbf{Parameter} & \textbf{Description} & \textbf{Prior} \\
			\hline
			\multicolumn{3}{c}{Merger rate evolution} \\[1pt]
			$R_0$ & PBH merger rate today in $\mathrm{Gpc}^{-3} \mathrm{yr}^{-1}$. & $\mU(0, 200)$\\
			\hline
			\multicolumn{3}{c}{Cosmological parameters} \\[1pt]
			\multirow{2}{*}{$H_0\,[\mathrm{km}\, \mathrm{s}^{-1} \mathrm{Mpc}^{-1}]$} & \multirow{2}{*}{Hubble constant.} & $\mU(10, 200)$ (Wide prior)\\
			& & $\mU(65, 77)$ (Restricted prior)\\
			\multirow{2}{*}{$\OM$} & \multirow{2}{*}{Present-day matter density of the Universe.} & $\mU(0, 1)$ (Wide prior)\\
			& & $\delta(0.315)$ (Restricted prior)\\
			\hline
			\multicolumn{3}{c}{Lognormal PBH mass function} \\[1pt]
			$\Mc\,[\Msun]$ & Peak mass of the lognormal mass function. & $\mU(5, 50)$\\
			$\sigma_\mathrm{c}$ & Mass width of the lognormal mass function. & $\mU(0.1, 2)$\\
			\hline
			\multicolumn{3}{c}{Power-law PBH mass function} \\[1pt]
			$\Mmin\,[\Msun]$ & Lower mass cut-off of the power-law mass function. & $\mU(3, 10)$\\
			\hline
			\multicolumn{3}{c}{Critical collapse (CC) PBH mass function} \\[1pt]
			$\Mf\,[\Msun]$ & Horizon mass scale of the CC mass function. & $\mU(5, 50)$\\
			$\al$ & Universal exponent of the CC mass function. & $\mU(0.5, 5)$\\
			\hline
			\multicolumn{3}{c}{ABH model} \\[1pt]
			$\gamma$ & Slope of the power-law regime for the rate evolution before the point $z_p$. & $\mU(0, 12)$\\
			$\kappa$ & Slope of the power-law regime for the rate evolution after the point $z_p$. & $\mU(0, 6)$\\
			$z_p$ & Redshift turning point between the powerlaw regimes with $\gamma$ and $\kappa$. & $\mU(0, 4)$\\
			$\alpha$ & Spectral index for the power-law of the primary mass distribution. & $\mU(1.5, 12)$\\
			$\beta$ & Spectral index for the power-law of the mass ratio distribution. & $\mU(-4, 12)$\\
			$m_{\min}\,[\Msun]$ & Minimum mass of the power-law component of the primary mass distribution. & $\mU(2, 10)$\\
			$m_{\max}\,[\Msun]$ & Maximum mass of the power-law component of the primary mass distribution. & $\mU(50, 200)$\\
			$\lambda_g$ & Fraction of the model in the Gaussian component. & $\mU(0, 1)$\\
			$\mu_g\,[\Msun]$ & Mean of the Gaussian component in the primary mass distribution. & $\mU(20, 50)$\\
			$\sigma_g\,[\Msun]$ & Width of the Gaussian component in the primary mass distribution. & $\mU(0.4, 10)$\\
			$\delta_m\,[\Msun]$ & Range of mass tapering at the lower end of the mass distribution. & $\mU(0, 10)$\\
			\hline
		\end{tabular}
	\end{adjustbox}
	\caption{\label{tab:priorss} Parameters and their prior distributions for the PBH scenario used in the Bayesian parameter estimations.}
\end{table*}

\begin{table}[tbp!]
	\centering
	\begin{tabular}{lc}
		\hline\hline
		PBH mass model\hspace{20mm} & $\log_{10}\mathcal{B}$ \\
		\hline
		Lognormal & $-0.02$ \\
		Power-law & $-0.11$ \\
		CC & $0.20$ \\
		\hline
	\end{tabular}
	\caption{The logarithm of the Bayes factor comparing runs that adopt the same PBH mass model but different cosmologies: Wide priors versus Restricted priors.}
	\label{tab:bf1}
\end{table}

\begin{table}[tbp!]
	\centering
	\begin{tabular}{ll}
		\hline\hline
		PBH mass model\hspace{20mm} & $\log_{10} \mathcal{B}$ \\
		\hline
		Lognormal & $2.99$ \\
		Power-law & $0$ \\
		CC & $3.12$ \\
		\hline
	\end{tabular}
	\caption{The logarithm of the Bayes factor between the different PBH mass models and the Power-law PBH mass model, for the case of a flat $\Lambda$CDM cosmology with wide priors.}
	\label{tab:bf2}
\end{table}

\begin{figure}[htbp!]
	\centering
	\includegraphics[width=.85\textwidth]{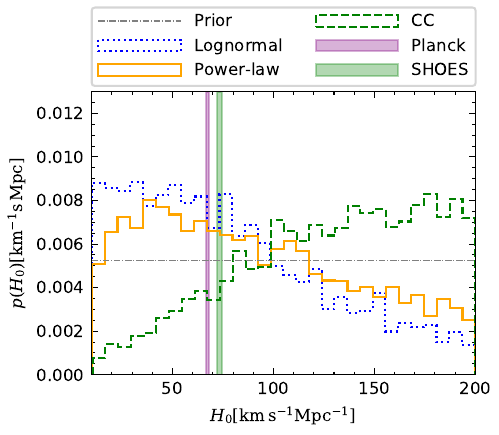}
	\includegraphics[width=.85\textwidth]{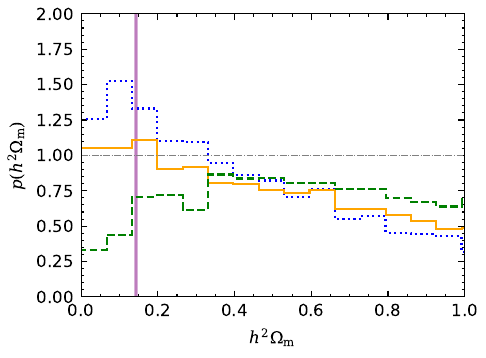}
	\caption{\label{H0_BBH} \textbf{Top panel}: One-dimensional marginal posterior distribution for $H_{0}$. \textbf{Bottom panel}: One-dimensional marginal posterior distribution for $h^2\Omega_{\mathrm{m}}$.
		In each panel, the blue dotted, orange solid, and green dashed lines represent the lognormal, power-law, and CC mass distributions of PBHs, respectively. The pink and green shaded areas indicate the $68 \%$ CI of the cosmological parameters inferred from CMB \citep{Planck:2018vyg} and the local Universe measurements \citep{Riess:2021jrx}, respectively.}
\end{figure}

In \Table{tab:bf1}, we report the Bayes factor between the general \LCDM\ model versus the $H_0$-tension model for three different PBH mass distributions, indicating no evidence of the data in favor of any one of these two cosmological models. This is mainly because $h^2\OM$ cannot be well constrained by the GW observations, and the uncertainty on the $H_0$ estimation extends far beyond the $H_0$ tension region, as can be seen from \Fig{H0_BBH}.

In \Table{tab:bf2}, we report the Bayes factors between different PBH mass models for the case of a general cosmology with wide priors. We find that the data strongly favor the lognormal and CC PBH mass models over the power-law model by a factor larger than $\sim 1000$, but no compelling evidence to prefer the lognormal PBH mass model over the CC mass model or vice versa.

\Fig{H0_BBH} shows the marginal posterior distributions for the cosmological parameters $H_{0}$ and $h^2\OM$ for the three different PBH mass models. The posteriors for these two cosmological parameters are broad and uninformative, indicating the current BBH events cannot constrain them, as anticipated by the Bayes factors discussed above.
\begin{figure}[htbp!]
	\centering
	\includegraphics[width=1\textwidth]{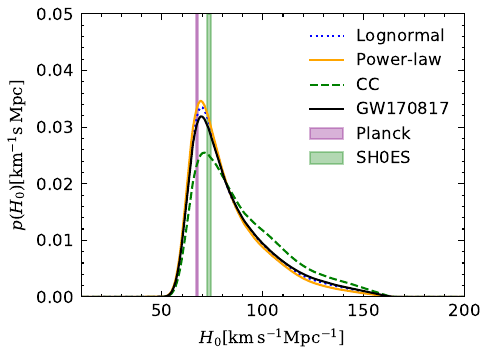}
	\caption{\label{H0_joint} The one-dimensional marginal posterior distribution for $H_0$ obtained by combining the $H_0$ posteriors from the $42$ BBH events and the $H_0$ posterior inferred from the bright standard siren GW170817. The blue dotted, orange solid, and green dashed lines represent the lognormal, power-law, and CC mass distributions of PBHs, respectively. The pink and green shaded areas indicate the $68 \%$ CI constraints on $H_0$ inferred from CMB \citep{Planck:2018vyg} and the local Universe measurements \citep{Riess:2021jrx}, respectively. The $H_0$ posteriors for GW170817 are adapted from \cite{LIGOScientific:2021aug}.}
\end{figure}
\Fig{H0_joint} shows the $H_{0}$ posteriors obtained by combining the $H_0$ posteriors from the three PBH mass models with the $H_{0}$ posteriors inferred from the bright standard siren GW170817 \cite{LIGOScientific:2017adf}. The combined estimation of the Hubble constant is $H_0 = 69^{+19}_{-8}$ \kmsmpc, $H_0 = 69^{+19}_{-8}$ \kmsmpc, and $H_0 = 70^{+26}_{-8}$ \kmsmpc, at the $68\%$ credible level for the lognormal, power-law, and CC PBH mass models, respectively.
Unless stated otherwise, credible intervals are quoted as maximum posterior and $68\%$ highest density intervals.
These results are at the same level compared with those obtained under the phenomenological mass models reported in \cite{LIGOScientific:2021aug}.

\begin{figure}[htbp!]
	\centering
	\includegraphics[width=1\textwidth]{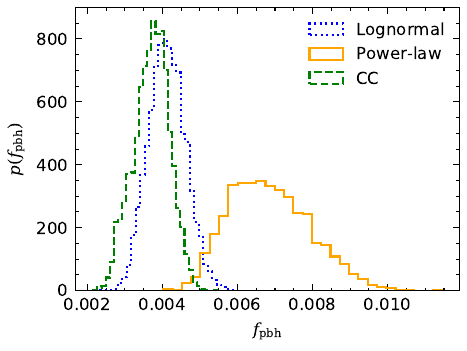}
	\caption{\label{fpbh} The one-dimensional marginal posterior distribution for the abundance of PBHs in CDM,  $\fpbh$. The blue dotted, orange solid, and green dashed lines represent the lognormal, power-law, and CC mass distributions of PBHs, respectively.}
\end{figure}

The local merger rate is $R_0 = 69^{+31}_{-22} \gpcyr$, $R_0 = 65^{+30}_{-21} \gpcyr$, and $R_0 = 93^{+37}_{-29} \gpcyr$ for the lognormal, power-law, and CC PBH mass distributions, respectively.
The corresponding posterior distributions for the $\fpbh$ parameter are shown in \Fig{fpbh}, with $\fpbh = 4.1^{+0.5}_{-0.8} \times 10^{-3}$, $\fpbh = 6.8^{+1.2}_{-1.0} \times 10^{-3}$, and $\fpbh = 3.7^{+0.4}_{-0.5} \times 10^{-3}$ for the lognormal, power-law, and CC PBH mass models, respectively.

\begin{figure}[htbp!]
	\centering
	\includegraphics[width=1\textwidth]{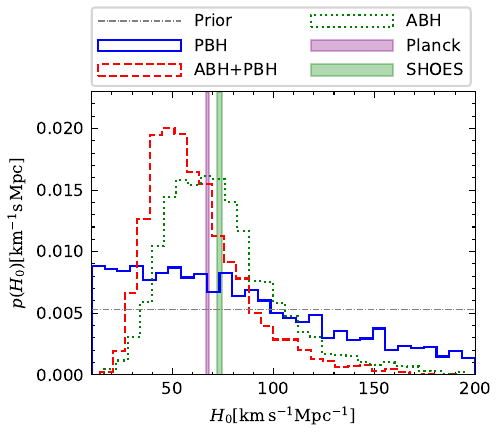}
	\caption{\label{H0_ABHPBH}One-dimensional marginal posterior distribution for $H_{0}$. The blue, red dashed, and green dotted lines represent the PBH, mixed ABH+PBH, and ABH models, respectively. The pink and green shaded areas indicate the $68 \%$ CI of the cosmological parameters inferred from CMB \citep{Planck:2018vyg} and the local Universe measurements \citep{Riess:2021jrx}, respectively.}
\end{figure}

For the mixed ABH+PBH model, we find that the fraction of detectable events of PBH binaries in the GWTC-3 is $f_{\rm P}\equiv N^{\rm det}_{\rm PBH}/(N^{\rm det}_{\rm PBH}+N^{\rm det}_{\rm ABH}) = 24.5^{+30.6}_{-17.3}\%$, consistent with \cite{DeLuca:2021wjr,Zheng:2022wqo}. Although the uncertainty on $f_{\rm P}$ is quite huge, this result implies that at least a few BBHs in GWTC-3 can be ascribed to the PBH channel.
In \Fig{H0_ABHPBH}, we show the posterior distributions for the Hubble constant derived from the single PBH, single ABH, and mixed ABH+PBH models. The measurements of $H_0$ are $H_0 = 70^{+62}_{-41}$ \kmsmpc, $H_0 = 57^{+27}_{-17}$ \kmsmpc  and $H_0 = 70^{+30}_{-21}$ \kmsmpc at the $68\%$ credible level for PBH, ABH+PBH and ABH models, respectively. It can be seen that the mixed ABH+PBH model can better constrain the Hubble constant than either the single ABH or PBH model.

\section{\label{conclusion}Discussion}

In this work, we constrain the Hubble parameter using the $42$ BBHs with detected $\mathrm{SNR}>11$ in the LVK GWTC-3 data release by assuming these BBHs are from PBHs. Three different PBH mass models are considered in the analyses. We find the data strongly disfavor the power-law PBH mass model by simultaneously inferring the population and the cosmological parameters. However, there is no compelling evidence to prefer the lognormal PBH mass model over the CC mass model or vice versa. The abundance of PBH in CDM, $\fpbh$, is at the order $\mathcal{O}(10^{-3})$ for all three PBH mass models, confirming that the stellar-mass PBHs cannot dominate CDM.

While the constraints on the present-day matter density, $\Omega_{\mathrm{m}}$ is weak and informative,
we estimate the Hubble constant to be $H_0 = 69^{+19}_{-8}\, \mathrm{km}\, \mathrm{s}^{-1}\, \mathrm{Mpc}^{-1}$ and $H_0 = 70^{+26}_{-8}\, \mathrm{km}\, \mathrm{s}^{-1}\, \mathrm{Mpc}^{-1}$ at $68\%$ confidence level for the lognormal and critical collapse mass functions, respectively, by combining the measurement from GW170817 and its EM counterpart \cite{LIGOScientific:2017adf}.
These results are at the same level as those obtained under the phenomenological mass models reported in Ref.~\cite{LIGOScientific:2021aug}.

Furthermore, using the mixed ABH+PBH model, we are able to get a more precise Hubble constant. This implies that a (small) fraction of PBHs in the total population would help decrease the uncertainty for the Hubble constant measurement and thus improve the constraint on the Hubble expansion. The more precise Hubble constant derived from the ABH+PBH scenario is expected because the Hubble constant highly degenerates with the mass distribution, and the mixed ABH+PBH model having more model parameters can better describe the BBH mass distribution in GWTC-3 than either one of the single ABH or single PBH model, thus helping to break the degeneracy between the Hubble constant and mass distributions. With the increased BBH events, the mixed ABH+PBH model provides a robust statistical inference for both the population and cosmological parameters.

\acknowledgments
We thank the referee for very useful comments.
We also thank Lang Liu, Xiao-Jin Liu, Zhu Yi, Xing-Jiang Zhu, and Zong-Hong Zhu for valuable discussions.
QGH is supported by the grants from NSFC (Grant No.~12250010, 11975019, 11991052, 12047503), Key Research Program of Frontier Sciences, CAS, Grant No.~ZDBS-LY-7009, CAS Project for Young Scientists in Basic Research YSBR-006, the Key Research Program of the Chinese Academy of Sciences (Grant No.~XDPB15).
ZCC is supported by the National Natural Science Foundation of China (Grant No.~12247176) and the China Postdoctoral Science Foundation Fellowship No.~2022M710429.
ZQY is supported by the China Postdoctoral Science Foundation Fellowship No.~2022M720482.

This research has made use of data or software obtained from the Gravitational Wave Open Science Center (gw-openscience.org), a service of LIGO Laboratory, the LIGO Scientific Collaboration, the Virgo Collaboration, and KAGRA. LIGO Laboratory and Advanced LIGO are funded by the United States National Science Foundation (NSF) as well as the Science and Technology Facilities Council (STFC) of the United Kingdom, the Max-Planck-Society (MPS), and the State of Niedersachsen/Germany for support of the construction of Advanced LIGO and construction and operation of the GEO600 detector. Additional support for Advanced LIGO was provided by the Australian Research Council. Virgo is funded, through the European Gravitational Observatory (EGO), by the French Centre National de Recherche Scientifique (CNRS), the Italian Istituto Nazionale di Fisica Nucleare (INFN) and the Dutch Nikhef, with contributions by institutions from Belgium, Germany, Greece, Hungary, Ireland, Japan, Monaco, Poland, Portugal, Spain. The construction and operation of KAGRA are funded by Ministry of Education, Culture, Sports, Science and Technology (MEXT), and Japan Society for the Promotion of Science (JSPS), National Research Foundation (NRF) and Ministry of Science and ICT (MSIT) in Korea, Academia Sinica (AS) and the Ministry of Science and Technology (MoST) in Taiwan.

\bibliographystyle{JHEP}
\bibliography{ref}

\newpage
\appendix
\section{\label{post}Full Posteriors for the PBH and Mixed ABH+PBH models}
This appendix shows the posteriors of all the cosmological and population parameters for the single PBH and the mixed ABH+PBH models considered in our analyses. The corner plots are produced using the \texttt{corner} \cite{corner} package.

\begin{figure*}[htbp!]
	\centering
	\includegraphics[width=\linewidth]{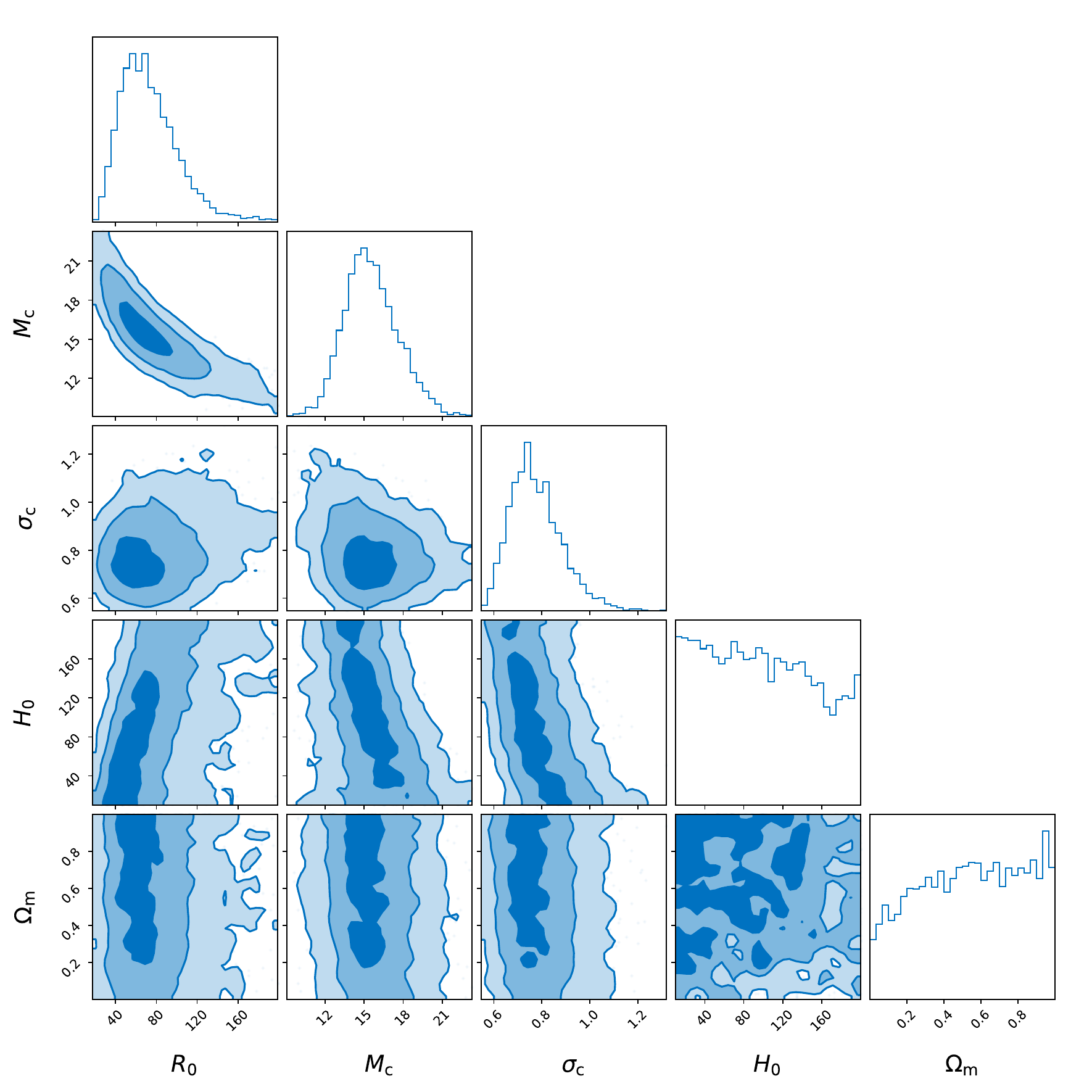}
	\caption{\label{lognormal-nocut-post}One and two-dimensional marginalized posteriors of the hyperparameters for the lognormal mass distribution in single PBH model. We show both the $1 \sigma$, $2 \sigma$, and $3 \sigma$ contours in the two-dimensional plot.}
\end{figure*}

\begin{figure*}[htbp!]
	\centering
	\includegraphics[width=\linewidth]{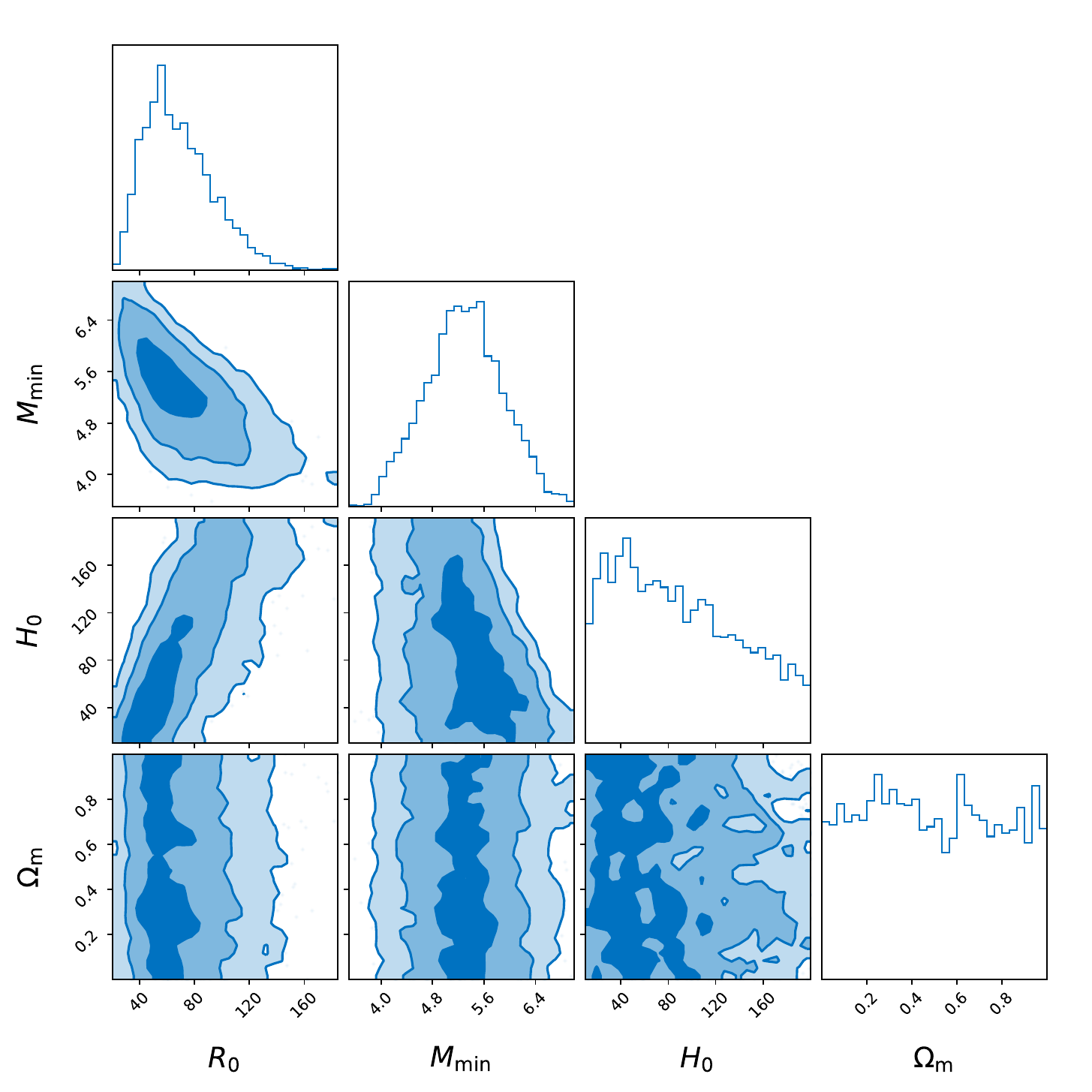}
	\caption{\label{powerlaw-nocut-post}One and two-dimensional marginalized posteriors of the hyperparameters for the power-law mass distribution in single PBH model. We show both the $1 \sigma$, $2 \sigma$, and $3 \sigma$ contours in the two-dimensional plot.}
\end{figure*}

\begin{figure*}[htbp!]
	\centering
	\includegraphics[width=\linewidth]{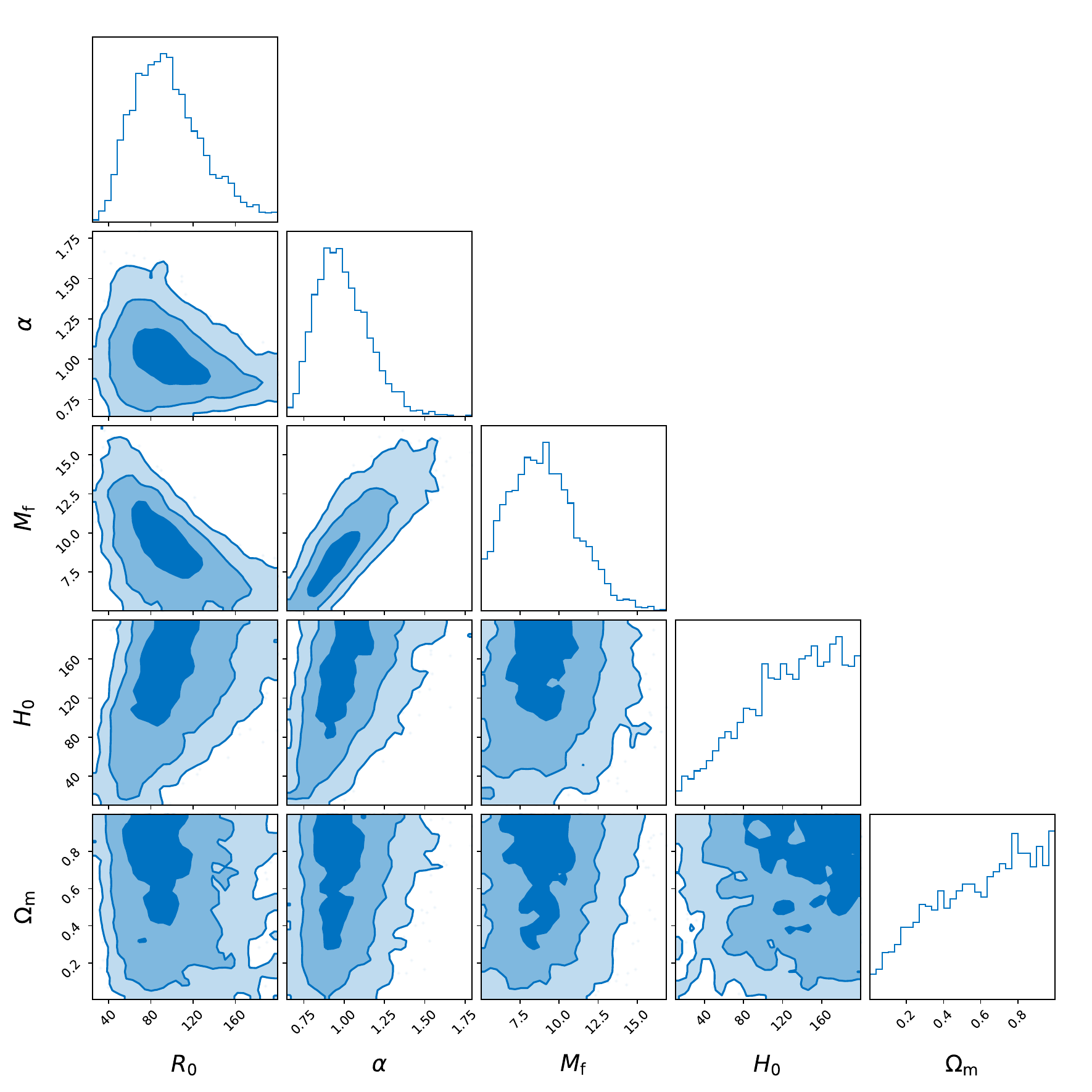}
	\caption{\label{CC-nocut-post}One and two-dimensional marginalized posteriors of the hyperparameters for the CC mass distribution in single PBH model. We show both the $1 \sigma$, $2 \sigma$, and $3 \sigma$ contours in the two-dimensional plot.}
\end{figure*}

\begin{figure*}[htbp!]
	\centering
	\includegraphics[width=\textwidth]{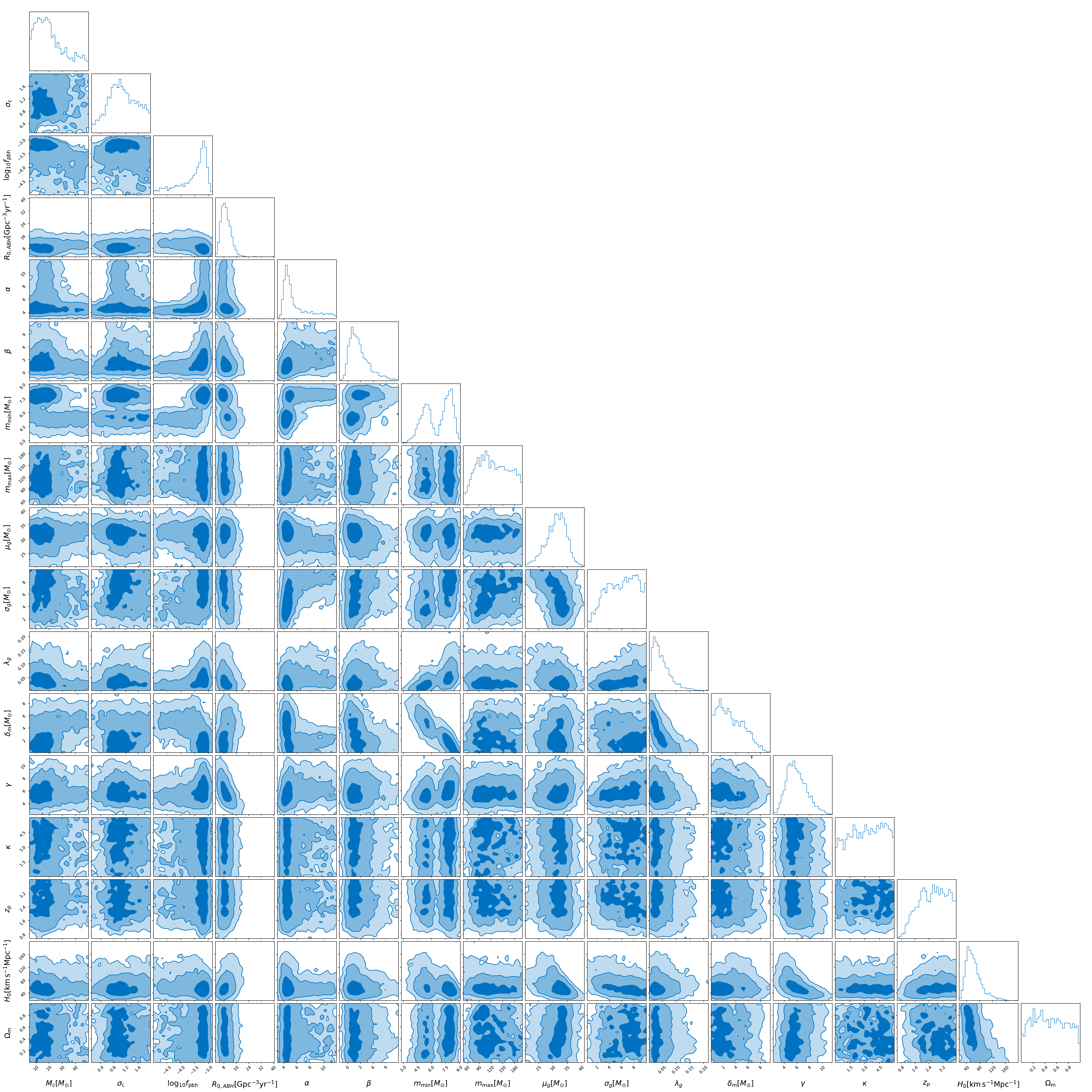}
	\caption{\label{abh-lognormal-nocut-post}One and two-dimensional marginalized posteriors of the hyperparameters for the mixed ABH+PBH model. We show both the $1 \sigma$, $2 \sigma$, and $3 \sigma$ contours in the two-dimensional plot.}
\end{figure*}

\end{document}